\documentclass{llncs}

\usepackage{fancyhdr}
\usepackage{flushend}
\usepackage{amsfonts,amssymb,amsmath,alltt}
\usepackage{stmaryrd}
\usepackage{xspace}
\usepackage{graphicx}
\usepackage{color}

\usepackage{makeidx}
\usepackage[utf8]{inputenc}
\usepackage{url}
\usepackage[english]{babel}

\newenvironment{ttbox}{\begin{alltt}\ttbraces\small\tt}%
                      {\end{alltt}}
\def\ttbraces{\let\.=\nobreak\chardef\{=`\{\chardef\}=`\}\chardef\|=`\\}

\newcommand\ttand{\mbox{{$\land$}}}
\newcommand\ttor{\mbox{{$\lor$}}}

\newcommand\ttcup{\mbox{{$\cup$}}}
\newcommand\ttfun{\mbox{{$\Rightarrow$}}}

\newcommand\ttimp{\mbox{{$\longrightarrow$}}}
\newcommand\ttequiv{\mbox{{$\equiv$}}}
\newcommand\ttexists{\mbox{{$\exists$}}}
\newcommand\ttforall{\mbox{{$\forall$}}}
\newcommand\ttneg{\mbox{{$\neg$}}}
\newcommand\ttneq{\mbox{{$\neq$}}}
\newcommand\ttin{\mbox{{$\in$}}}
\newcommand\ttnin{\mbox{{$\notin$}}}
\newcommand\ttImp{\mbox{{$\Longrightarrow$}}}
\newcommand\ttlam{\mbox{\( \lambda \)}}
\newcommand\tttimes{\mbox{\( \times \)}}

\newcommand\ttatI{\mbox{\( @_G \)}}

\newcommand\ttrelIstar{\mbox{{$\to^*$}}}
\newcommand\ttrel[1]{\mbox{{$\to_{#1}$}}}

\newcommand\ttvdash{\mbox{{$\vdash$}}}

\newcommand{\ttcalN}[1]{\mbox{{${\mathcal{N}}_{\texttt{#1}}$}}} 

\newcommand\ttattand[1]{\mbox{{$\oplus_{\wedge}^{#1}$}}}

\begin{document}
\frontmatter
  
\mainmatter
\title{Exploring rationality of self awareness in social networking for logical modeling of unintentional insiders
}
\author{Florian Kamm\"uller and Chelsea Mira Alvarado}

\institute{Middlesex University London and\\ Technische Universit\"at Berlin\\
\email{f.kammueller@mdx.ac.uk|CA936@live.mdx.ac.uk}
}
\maketitle
\begin{abstract}
  Unawareness of privacy risks together with approval seeking motivations make humans enter too
  much detail into the likes of Facebook, Twitter, and Instagram. To test whether the rationality
  principle applies, we construct a tool that shows to a user what is known publicly on social
  networking sites about her. In our experiment, we check whether this revelation changes human behaviour.
  To extrapolate and generalize, we use the insights gained by practical experimentation. Unaware users
  can become targeted by attackers.  They then become unintentional insiders. We demonstrate this by extending
  the Isabelle Insider framework to accommodate a formal model of unintentional insiders, an open problem with
  long standing.
\end{abstract}

\section{Introduction}
\label{sec:intro}
  The privacy paradox \cite{Barnes_2006} shows that humans can be on one side quite concerned about security
  and privacy in general but when it comes to their own behaviour they seem to ignore any caution and freely
  spread their private data into public cloud based social network services, like Facebook, Twitter or
  Instagram. 
  Assuming that humans are rationally acting beings has led to quite successful models and prediction in
  economics using what is termed Rational Choice Theory (RCT) \cite{sco:00}. Sociologist have transferred RCT
  more generally to social interaction forming what is known  as {\it exchange theory}.
  We want to test this theory on the privacy paradox and use the results to improve automated
  logical verification of social networks. We consider a dynamic system research approach more suitable.
  The Isabelle Insider framework permits modeling and analyzing dynamic state transition. Thus we can reason
  on actions and their effects. Methodologically, we thus follow the action research approach \cite{lew:48}
  interleaving empirical research with interventions, here, practical implementations and verification.
  
  This paper first presents an empirical study on increasing privacy awareness for
  the construction of a social self awareness tool for social networks.
  It uses assumptions
  from RCT  testing and highlighting the significance
  of applying this theory.
  RCT can be considered as a follow up theory of Max Weber's sociological explanation which has
  strongly inspired the human actor model of Isabelle's Insider framework. Consequently, it appears
  natural to use the RCT interpretation found in the empirical study to extend the human model in the
  Isabelle Insider framework. 
  Moreover, it turns out that the RCT interpretation of social awareness allows to model unintentional
  insiders a challenge hitherto unanswered.

  This paper first gives some background from sociology about RCT and the Isabelle Insider framework
  (Section \ref{sec:back}).
  Section \ref{sec:snpa} presents the tool based study on privacy awareness in social networks and the influence
  of RCT giving some insights into the requirements, design, testing and evaluation of the tool and the
  key findings in the RCT interpretation with respect to privacy awareness. This section is based on the
  Bachelor of Science dissertation of one of the authors \cite{alv:21}.
  Section \ref{sec:isaunins} then continues to transfer the experimental findings into extending
  the Isabelle Insider framework and illustrating them on the case study. The Isabelle sources are
  publicly available on github \cite{kam:21}.

\section{Background}
\label{sec:back}
\subsection{Social Explanation and Rational Choice Theory}
\begin{figure}
  \begin{center}
  \includegraphics[scale=.5]{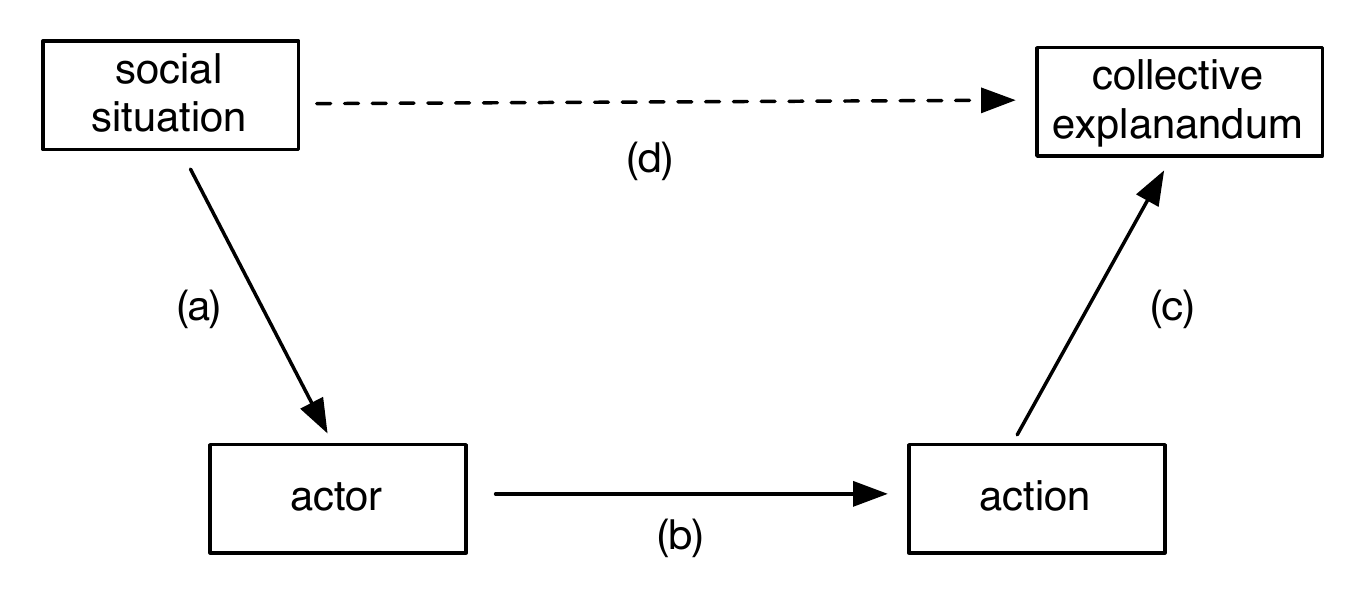}
  \caption{Max Weber's sociological explanation model: a macro-micro-macro-level-transition explaining 
sociological phenomena by breaking down the global facts from the macro level (a) onto a more 
refined local view of individual actors at the micro-level (b). Finally those micro-steps
are generalized and lifted back on the macro-level (c) to explain the global phenomenon (d).}\label{fig:weber}
  \end{center}
\end{figure}

Rational choice theory is based upon the assumption that  complex
social phenomena can be explained by individual actions that constitute them. This
philosophy now coined {\it methodological individualism} holds that:
`The elementary unit of social life is the individual human action. To explain social institutions
and social change is to show how they arise as the result of the action and interaction of individuals'
\cite{els:89}.
Seemingly very close to {\it methodological individualism}, is 
what was originally conceived by Max Weber \cite{we:72,we:20a} as
`understanding explanation' ({\it Verstehendes Erkl\"aren}) sketched in Figure \ref{fig:weber}.
Despite these similarities, RCT is more extreme in only considering rational actions.
John Scott explains in his critical overview over RCT \cite{sco:00}:
`what distinguishes RCT [\dots] is that it denies the existence of any kind of action
other than the purely rational [\dots]'. We draw from Scott's overview to 
contrast and provide the right context for our work.
He quite critically highlights the limitations of RCT in particular when branching out
from economics and applying RCT more generally to sociology. According to Scott, Homans
\cite{hom:61} was a ``pioneering figure'' in establishing rational choice theory in sociology
setting up the basic framework of {\it exchange theory} which can be understood as RCT for social
interaction. In this framework, money and market mechanisms of economic theories are
replaced by human resources as time, information, approval and prestige.
Besides  pioneering RCT, Homans additionally grounded exchange theory on assumptions that
he drew from behaviourist psychology.
While the methodological individualism of rational choice theories starts from individuals' actions
and sees all social phenomena reducible to these actions, Homans went one step further into
explaining them. For him it was necessary to reduce these actions to conditioned psychological responses.
In brief, human behaviour is like animal behaviour not free but determined by rewards and
punishment. This reinforcement is called `conditioning' and determines human behaviour.
Behaviour can thus be studied purely externally and needs no inspection of internal mental
states.

While others rejected Homans' claims about this explanation of human behaviour
-- and even Homans came to see it as inessential -- for our formal model of awareness and
unintentional insiders it is very helpful. In Section \ref{sec:isaunins}, when we formalize
the taxonomy extracted from the experimental work into Isabelle, we model human behaviour
in the sense of `conditioning'. We actually do model the internal state of the actors
although Homans considered this as unnecessary but our model permits dynamic state inspection including
psychological disposition of human actors.

\subsection{Isabelle Insider framework}
The Isabelle Insider framework \cite{bikp:14,kp:16} has also been inspired by Max Weber and
methodological individualism.
In mapping this fundamental philosophy to logic, this framework follows a common introductory
textbook for sociologists by Hartmut Esser \cite{he:93} written in the spirit of Popper's
critical rationalism. This offers an approach to understand sociological experiments in a
formal way using a logical view on explanation by the logicians Hempel and Oppenheim \cite{ho:48}.
In addition, the Isabelle Insider framework uses a taxonomy provided in \cite{nblgcww:14}
which is founded on empirical and psychological studies of counterproductive workplace behaviour.
In Section \ref{sec:isaunins}, we will in more detail present the details of how the human disposition
and its effects to the environment are modeled in Isabelle and how this model is now extended to accommodate
the unintentional insider.

Isabelle is an interactive proof assistant based on Higher Order Logic (HOL). 
Application specific logics are formalized into new theories extending HOL.
They are called object-logics. Although HOL is undecidable and therefore proving
needs human interaction, the reasoning capabilities are very sophisticated
supporting ``simple'', i.e., repetitive, tedious proof tasks to a level of
complete automation. The use of HOL has the advantage that it enables expressing
even the most complex application scenarios, conditions, and logical
requirements and HOL simultaneously enables the analysis of the meta-theory. 
That is, repeating patterns specific to an application can be abstracted and
proved once and for all.
An object-logic contains new types, constants, and definitions. These items 
reside in a theory file. For instance, the file \texttt{UnintentionalInsider.thy} contains
the object-logic for unintentional insiders described in the following paragraphs. 
This Isabelle Insider framework is a {\it conservative extension} of HOL. This means
that our object logic does not introduce new axioms and hence guarantees consistency.
Conceptually, new types are defined as subsets of existing types and properties are proved
using a one-to-one relationship to the new type from properties of the existing type.

We are going to use Isabelle syntax and concepts in the presentation of the Isabelle Insider
framework and will explain them when they are used.

\section{Social Networks and Privacy Awareness}
\label{sec:snpa}
\subsection{Requirements analysis and design of social awareness tool}
A questionnaire was created in order to research about public attitudes to internet security
amongst social media users. Quantitative and qualitative data are obtained through this  method,
allowing more time for analysis of the results and how the results can be used to create a
prototype.
Answers to `How many different social media apps/websites do you use every day?',
show that 84.6\% of 39 responses use more than 3 different forms of social media every day. This
shows the commonality and reliance of social media in everyday lives and how many different apps
can hold information about you.

\includegraphics[scale=.3]{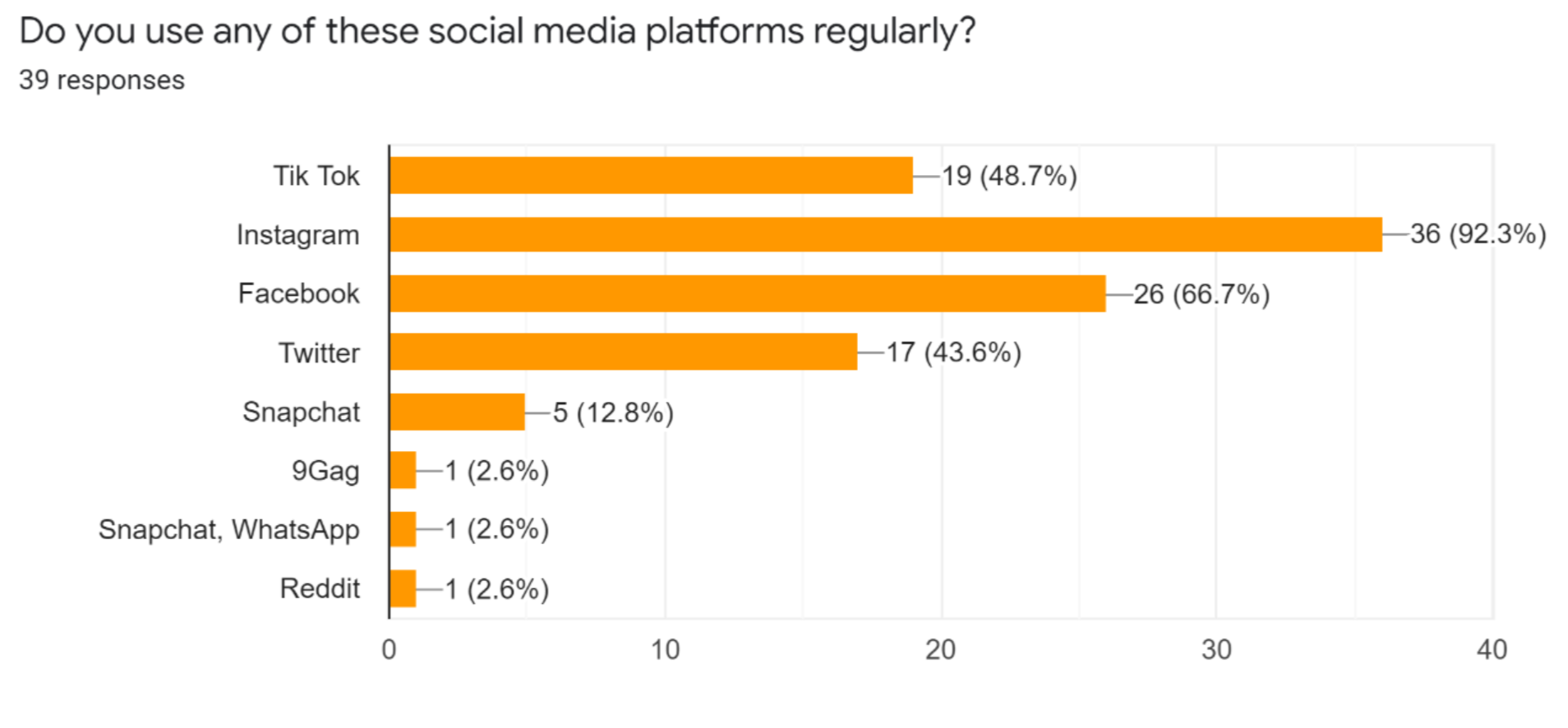}

The above chart show the most common applications that people use, Instagram, Facebook and TikTok
have shown to be the most common platforms. Each of which have shown to have breaches in misuse of
personal data they have collected from their users.

How many people have their accounts on private? The majority, 56.4\%, say that only some are private,
meaning that the users have chosen to only privatise one or more of their accounts but have left others
to be able to be accessed by the public.
Are these users aware of how much information they have put out publicly? Surprisingly, the most common
answers are completely aware or somewhat aware. Of the amount. 59\% have said that they have knowledge
of information they have posted publicly but leave room for uncertainty as to how much is actually available
to the public. This shows a slight concern from users in their social media behaviour.

\subsection{Testing and Evaluation}
\begin{figure}
  \begin{center}
    \includegraphics[scale=.25]{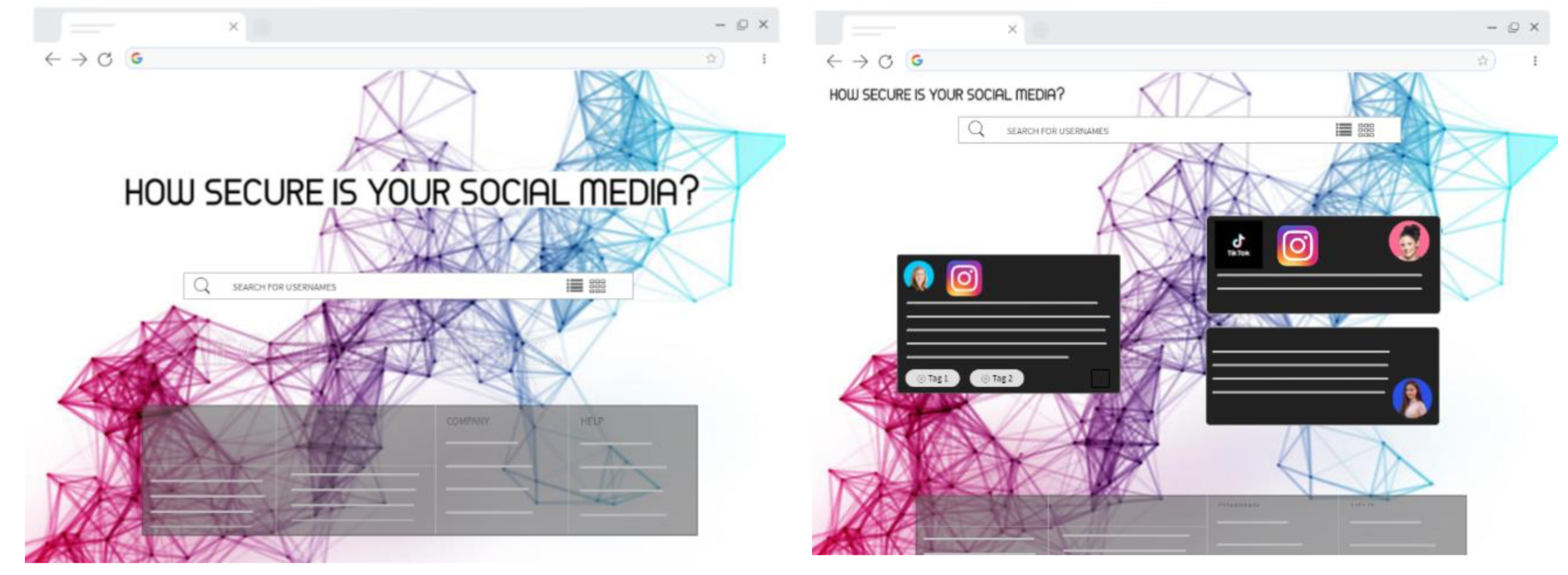}
    \end{center}\caption{Interface of social network awareness tool}\label{fig:interface}
\end{figure}

The left side of Figure \ref{fig:interface} shows the design of the search page which focuses on clear
minimalistic esthetic to display clear concise information, which will be easily accessible by all. The
website title placed at the top middle and highlights the purpose pf the website. The search bar is in the
middle of the page, letting the user know that the tool only has one importance and should not show otherwise.
The user will not be lost when navigating the website, easing user comfort. The bottom grey section reflects
basic information on the importance of internet security and what the website aims to show. User inputs through
the search bar and uses the search button.
The right side of the figure displays what the user sees when they have searched a username. Profiles are created
of the available information such as other social media accounts that are linked to the username searched.
The profiles are highlighted by the black boxes they are in that contrast the white background, allowing less
crowded visuals which may have disorientated people.

For the implementation, we used opensource API’s. An API is an application programming interface that allows
computers to send signals and receive data in return. This enables specific queries and actions to be retrieved.
APIs need keys allowing access to sensitive data whilst also protecting important and sensitive data that
cannot be accessed by any user. All social networks allow developers to apply for API keys, allowing APIs to be
used for projects. The API allowed us to retrieve that data necessary enabling to connect to the internet and
use genuine  social network server data. Users are able to search any username on any social
network and retrieve related information. The API’s proved to be the best solution for this project as
we could acquire the necessary data and use it to create a summarized only profile.
The results are thus inherently genuine reflecting real world scenarios.


\subsection{Key findings and RCT interpretation of privacy awareness}
\includegraphics[scale=.3]{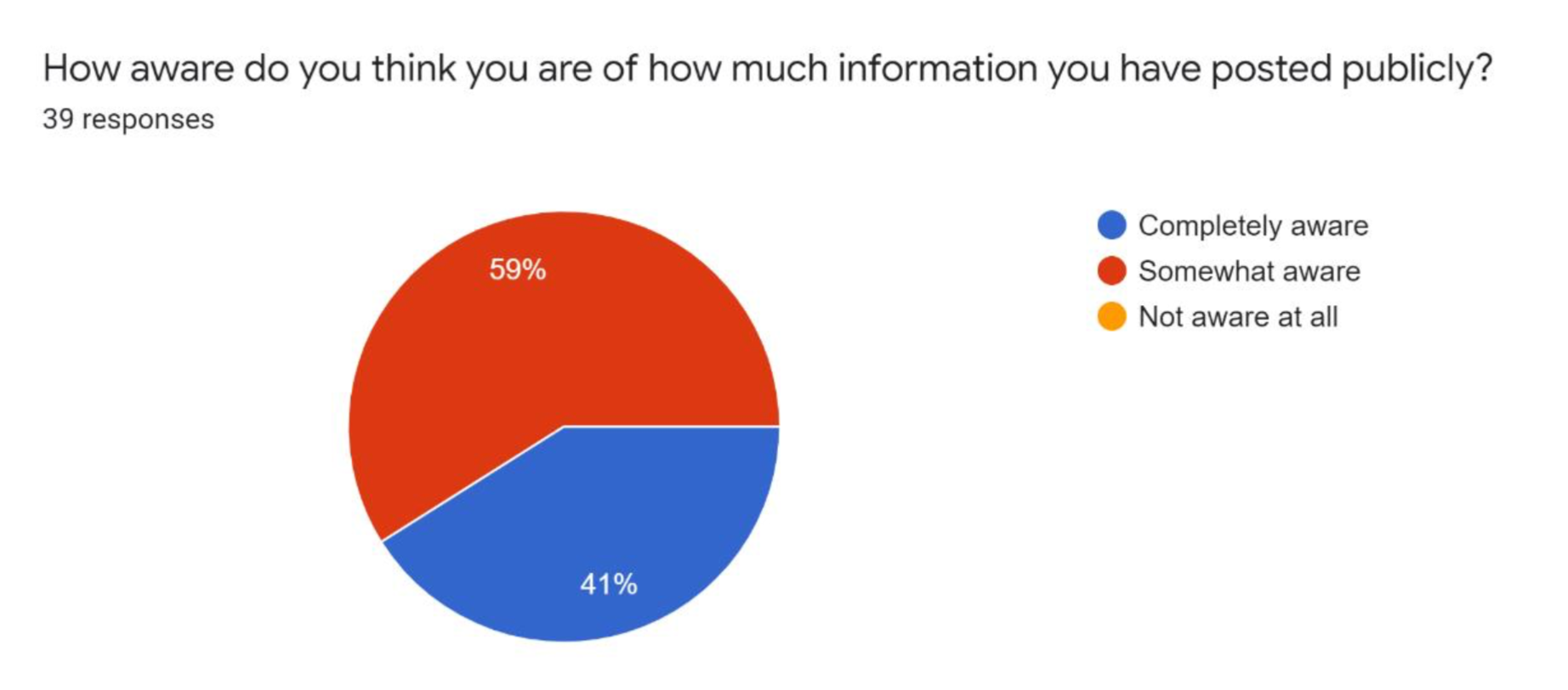}

Most importantly, the last question investigates the issue of social media behaviours. It asks whether they
would personally change their online behaviours if they were able to see what others could see about them.
From responses to this question on the questionnaire, many have shown concern about what others could obtain
from the information they post online and would immediately act on this by privatising their social media.
Consequently, this shows the importance of a tool that helps people become more aware of their online behaviours.

It matches the rationality assumption of RCT and proves that creating awareness changes the users attitude.
This creates a potential for improved privacy in social networks and how awareness could change the risk of
attacks on privacy.

\section{Modeling Unaware Social Network Users and Unintentional Insiders in Isabelle}
\label{sec:isaunins}
The state based dynamic semantics of the Isabelle infrastructure framework allows expressing how
awareness dynamically changes the global policy and thus how a change on awareness eliminates the risk.
We also show how to integrate awareness into the notion of insiderness thus extending the Isabelle
Insider framework to unintentional insiders based on the findings of our experiment with the social
awareness tool.
\subsection{Infrastructures, Policies, Actors in Isabelle}
\label{sec:infra}
The Isabelle Infrastructure framework supports the representation of infrastructures 
as graphs with actors and policies attached to nodes. These infrastructures 
are the {\it states} of the Kripke structure. 

The transition between states is triggered by non-parameterized
actions \texttt{get}, \texttt{move}, \texttt{eval}, and \texttt{put} 
executed by actors. 
Actors are given by an abstract type \texttt{actor} and a function 
\texttt{Actor} that creates elements of that type from identities 
(of type \texttt{string} written \texttt{''s''} in Isabelle). 
\begin{ttbox}
{\bf {typedecl}} actor 
{\bf {type}}_{\bf{synonym}} identity = string
{\bf {consts}} Actor :: string \ttfun actor
\end{ttbox}
Note that it would seem more natural and simpler to just 
define \texttt{actor} as a datatype over identities with a constructor \texttt{Actor}
instead of a simple constant together with a type declaration like, for example,
in the Isabelle inductive package by Paulson~\cite{pau:97}. 
This would, however, make the constructor \texttt{Actor} an injective function
by the underlying foundation of datatypes therefore excluding the fine
grained modelling that is at the core of the insider definition:
in fact, {the core insider property \texttt{UasI} (see below)} defines the
function \texttt{Actor} to be injective for all except insiders and explicitly enables
insiders to have different roles by identifying \texttt{Actor} images.

To represent the macro level view seeing the actor within an infrastructure,
we define a graph datatype \texttt{igraph} (see below) for infrastructures.
This datatype has generic input parameters that are going to be supplied as
concrete parts of an application infrastructure on instantiation of an \texttt{igraph}.
They represent the actual location graph, the actors in each locations, their roles, credentials
and psychological disposition (see following subsection) and the locations' state.
\begin{ttbox}
{\bf datatype} igraph = Lgraph (location \tttimes location)set 
                                location \ttfun identity set
                                actor \ttfun (string list \tttimes string list)
                                actor \ttfun actor_state
                                location \ttfun string list
\end{ttbox}
Consider here the social network case study as an example.
\begin{ttbox}
 ex_graph \ttequiv  Lgraph 
  \{(aphone,instagram), (bphone,instagram)\}
    (\ttlam x. if x = aphone then \{''Alice''\} else 
      (if x = bphone then \{''Bob''\} else \{\})) 
      ex_creds ex_locs
\end{ttbox}
Policies specify the expected behaviour of actors of an infrastructure. 
Atomic policies of type \texttt{apolicy}
describe prerequisites for actions to be granted to actors given by
pairs of predicates (conditions) and sets of (enabled) actions:
\begin{ttbox}
{\bf type}_{\bf{synonym}} apolicy = ((actor \ttfun bool) \tttimes action set)
\end{ttbox}
For example, the \texttt{apolicy} pair
\texttt{(\ttlam x.\ {has (x, ’’PIN’’)}, \{move\})} 
specifies that all actors {who know the \texttt{PIN}} are enabled to perform action \texttt{move}.

Infrastructures combine an infrastructure graph of type \texttt{igraph}
with a policy function that assigns local policies 
over a graph to each location of the graph, that is, it is a function
mapping an \texttt{igraph} to a function from \texttt{location} to 
\texttt{apolicy set}.

\begin{ttbox}
{\bf datatype} infrastructure = Infrastructure igraph 
                                          [igraph, location] \ttfun apolicy set
\end{ttbox}
For our social network example, the initial infrastructure contains the above graph
\texttt{ex\_graph} and the local policies defined shortly. 
\begin{ttbox}
sn_scenario \ttequiv Infrastructure ex_graph local_policies
\end{ttbox}
The function \texttt{local\_policies} gives the policy for each location \texttt{x} 
over an infrastructure graph \texttt{G} as a pair: the first element of this pair is 
a function specifying the actors \texttt{y} that are entitled to perform the actions 
specified in the set which is the second element of that pair.
\begin{ttbox}
 local_policies G x \ttequiv
 case x of 
   aphone \ttfun \{(\ttlam y. has G (y,''aPIN'')), \{put,get,move,eval\})\}
 | bphone \ttfun \{((\ttlam y. has G (y,''bPIN'')), \{put,get,move,eval\})\} 
 | instagram \ttfun \{(\ttlam y. \ttin \{Actor ''Alice'', Actor ''Bob''\},
                          \{put,get,move,eval\})\}
 | _ \ttfun  \{\})
\end{ttbox}
We define the behaviour of actors using a predicate \texttt{enables}:
within infrastructure \texttt{I}, at location \texttt{l},
an actor \texttt{h} is enabled to perform an action \texttt{a} if there
is a pair \texttt{(p,e)} in the local policy of \texttt{l} --  \texttt{delta I l} 
projects to the local policy -- such that action \texttt{a} is in the action set 
\texttt{e} and the policy predicate \texttt{p} holds for actor \texttt{h}.
\begin{ttbox}
enables I l h a \ttequiv \ttexists (p,e) \ttin delta I l. a \ttin e \ttand p h
\end{ttbox} 
For example, the statement 
\texttt{enables I l (Actor''Bob'') move} is true if 
the atomic policy \texttt{(\ttlam x. True, \{move\})} is in the set
of atomic policies \texttt{delta I l} at location \texttt{l} in infrastructure
\texttt{I}. {Double quotes as in \texttt{''Bob''} create a string in Isabelle/HOL.}

\subsection{Modelling the human actor and psychological disposition}
\label{sec:human}
The human actor's level is modeled in the Isabelle Insider framework by assigning
the individual actor's psychological disposition to each actor's identity.
\begin{ttbox}
{\bf datatype} actor_state = State psy_state motivations
\end{ttbox}
There are selector functions \texttt{motivation} and \texttt{psy\_state} to project
the components from an \texttt{actor\_state} element.
The psychological state of an actor is not determined using the formal system but we
use here empirical facts as input as for example our own 
studies from Section \ref{sec:snpa} or other sociological findings, like \cite{nblgcww:14}.
The formal representation of {\it Psychological State} 
is a simple enumeration datatype distinguishing the ``normal'' state of happiness from one in which the actor is
alerted or ``suspicious''.
\begin{ttbox}
{\bf datatype} psy_states = happy | suspicious
\end{ttbox}
The element on the right hand side are the two injective constructors of the new datatype
\texttt{psy\_states}. They are simple constants, modeled as functions without arguments.
Motivation plays a vital role in RCT and as Homans observed the strongest one is that humans
seek approval (which is only excluded by a state of mind that corresponds to complete detachment
which we abbreviate as ``zen'').
\begin{ttbox}
{\bf datatype} motivations = approval_hungry | zen
\end{ttbox}
The types for psychological state and motivations allow defining the users
state of unawareness by a predicate.
\begin{ttbox}
{\bf definition} unaware :: actor\_state \ttfun bool
  unaware a \ttequiv motivation a = \{approval_hungry\} \ttand happy = psy_state a 
\end{ttbox}

\subsection{Privacy by labeling data and state transition}
\label{sec:label}
The Decentralized Label Model (DLM) \cite{ml:98} introduced the idea to
label data by owners and readers. We use this idea and formalize 
a new type to encode the owner and the set of readers of a data item.
\begin{ttbox}
{\bf type\_synonym} dlm = actor \tttimes actor set
\end{ttbox}
Labelled data is then just given by the type \texttt{dlm \tttimes\ data}
where \texttt{data} can be any data type.

The abstract state transition provided in the underlying Kripke structure
theory is instantiated in the infrastructure model by an inductive definition 
of a state transition relation \texttt{\ttrel{n}} over infrastructures. 
A set of inductive rules defines this transition relation \texttt{\ttrel{n}} 
relative to characteristics of the current state. These characteristics can exploit
the information encoded into the infrastructure as well as the enables predicate
to express how the next infrastructure state evolves from the current one.
We show here the rules for put and get as they suffice to illustrate how to model
the social network application scenario.

\subsubsection{The put pata rule} assumes an 
actor \texttt{h} residing at a location \texttt{l} in the infrastructure
graph \texttt{G} and being enabled the \texttt{put} action. In addition, the psychological
state \texttt{pgra G h} needs to be unaware. Here we add the newly extended option for the
human actor model to the semantic rule as precondition thus stating that only unaware users
put their data onto the graph.
If infrastructure state \texttt{I} fulfills those preconditions, the next state \texttt{I'} can
be constructed from the current state by adding the data item 
\texttt{((Actor h, hs), n)} at location \texttt{l}. The addition is
given by updating (using \texttt{:=}) the existing data storage \texttt{lgra G l} 
at location \texttt{l} with the singleton set \texttt{\{((Actor h, hs), n)\}}. Note
that the first component \texttt{Actor h} marks the owner of this data item as \texttt{h}.
\begin{ttbox}
put: 
G = graphI I \ttImp h \ttatI l \ttImp enables I l (Actor h) put \ttImp
unaware (pgra G h) \ttImp
I' = Infrastructure 
       (Lgraph (gra G)(agra G)(cgra G)(pgra G)
         ((lgra G)(l := (lgra G l \ttcup \{((Actor h, hs), n)\}))))
       (delta I) 
\ttImp I \ttrel{n} I' 
\end{ttbox}

\subsubsection{The get data rule} resembles the put data rule in 
many parts. However, here an actor \texttt{h} accesses data in a remote
location \texttt{l'} and adds it to the data in his current location  
\texttt{l}. This copying of data is only permitted if the current location
\texttt{l'}  of the data enables \texttt{h} to \texttt{get} and if the list of readers
\texttt{hs} in the data item \texttt{((Actor h', hs), n)} contains the entry 
\texttt{Actor h} or if the accessing actor is \texttt{h} herself.
\begin{ttbox}
get_data: 
G = graphI I \ttImp h \ttatI l \ttImp enables I l' (Actor h) get \ttImp
((Actor h', hs), n) \ttin lgra G l' \ttImp Actor h \ttin hs \ttor h = h' \ttImp
I' = Infrastructure 
       (Lgraph (gra G)(agra G)(cgra G)(pgra G)
          ((lgra G)(l := (lgra G l \ttcup \{((Actor h', hs), n)\}))))
       (delta I) 
\ttImp I \ttrel{n} I' 
\end{ttbox}
The global policy is `only the owner and friends can access the data on the cloud'
using for example the definition of \texttt{friends} as \texttt{\{''Alice'', ''Bob''\}}.
\begin{ttbox}
 global_policy I a \ttequiv  a \ttnin friends
               \ttimp \ttneg(enables I instagram (Actor a) get)
\end{ttbox}
We can prove that Bob is enabled to get (Alice's data) at instagram if Bob is specified
as a reader in an application scenario where Alice sets the label parameter \texttt{hs}
in a put action accordingly. So, using the features of attack tree analysis of the
Isabelle Insider framework, we can formally prove such statements.
However, we are interested in  investigating negative effects of unawareness and how a change of human
behaviour may improve the situation. Therefore, we use the representation of human factors and
(malicious) Insiders in the Isabelle Insider framework, integrating the existing notion of
malicious insiders and extending them to include also unintentional insiders.

\subsection{Representing human factors and insiders}
The Isabelle Insider framework defines ``[a]n insider [as] a trusted user of a system who behaves
like an attacker abusing privileges thereby bypassing security controls'' \cite{kk:21}.
This definition leads to the notion of an insider as 
an attacker 
formally represented as an actor \texttt{Eve} who is a malicious ``evil'' actor outside some
set of actors within the system. Actors are represented as having a unique identity as well as a role of actor
which normally is the same as their identity unless impersonation happens. Insiderness is now represented
by explicitly identifying the actor \texttt{Eve} with privileged users. Thus the malicious actor Eve can
act like an inside actor.
So far, the Isabelle Insider framework has rooted insiderness on a taxonomy from the insider
threat literature based on psychological studies \cite{nblgcww:14}. Thus, insiderness was uniquely
determined by the description of an insider as a system actor turning bad as a consequence of susceptible
dispositions and triggering events leading to a ``tipping point''.

Technically, we model this explicit yet flexible impersonation of privileged users inside the system
by a function \texttt{Actor} that maps identities to roles. In places where an impersonation is deemed
feasible the function may map the identity of the ``evil'' actor \texttt{Eve} to the same role as
that of a privileged user inside the system. 
For all other identities that are not compromised the
function actor maps these identities exclusively to roles in the system, that is, for these
identities \texttt{Actor} is injective: $id_0 \neq id_1 \Rightarrow \texttt{Actor}\ id_0 \neq \texttt{Actor}\ id_1\,.$

Here, we want to extend this classical view of an intentional insider to that of an unintentional 
insider \cite{InsiderUnintentionalInsider2013}. 
As Matt Bishop puts it ``[i]n many cases, unintentional insider attacks are as dangerous as deliberate
insider attacks; preventing them adds more complexity to an already, difficult problem. Any approach
therefore must have not only a technical aspect (detecting the attack), but also a non-technical
aspect (detecting the problem), which includes consideration of social, political, legal, and cultural
influences, among others''\cite{Bishop:2017aa}.

We remain in the spirit of this design decision of representing the human actor but extend it with awareness
and thus unintentional insiderness. In the following we retrace the steps of the formal insider model
as originally conceived in the Isabelle Insider framework highlighting the additions and extensions
to accommodate unintentional insiders.

\subsection{Integrating Unaware with Malicious Insiders}
For the integration of unintentional insiders with the existing the malicious insiders, e.g. \cite{kk:21},
we extend the definitions of the types \texttt{motivations} and \texttt{psy\_state} given in Section \ref{sec:human}. 
The values for the malicious insider 
are based on a taxonomy from psychological insider research by Nurse et al.~\cite{nblgcww:14}.
\begin{ttbox}
{\bf datatype} psy_states = ... | depressed | disgruntled | angry | stressed
\end{ttbox}
Another example is
{\it motivation} for malicious insiders ranging far \cite{nblgcww:14}.
\begin{ttbox}
{\bf datatype} motivations = ... | financial | political | revenge 
                 | fun | competitive_advantage | power | peer_recognition
\end{ttbox}
The transition 
to become an insider is represented by a {\it catalyst} that tips the insider 
over the edge so he acts as an insider formalized as a ``tipping point''
predicate.
\begin{ttbox}
{\bf definition} tipping_point :: actor\_state \ttfun bool
\ tipping_point a \ttequiv motivation a \ttneq \{\} \ttand motivation a \ttneq \{approval_hungry\} 
                   \ttand  happy \ttneq psy_states a 
\end{ttbox}
To embed the fact that the attacker is an insider, the actor can then
impersonate other actors.
This assumption entails that an insider \texttt{Actor ''Eve''} can act like 
their alter ego, say \texttt{Actor ''Charlie''} within the context of the locale.
This is realized by the predicate  \texttt{UasI}.
\begin{ttbox}
UasI a b \ttequiv (Actor a = Actor b) \ttand
           \ttforall x y. x \ttneq a \ttand y \ttneq a \ttand Actor x = Actor y \ttimp x = y
\end{ttbox}
Note that this predicate also stipulates that the function \texttt{Actor} 
is injective for any other than the identities \texttt{a} and \texttt{b}.
This completes the Actor function to an ``almost everywhere injective function''.
Insiderness can now be defined as a rule that is triggered by conditions that may
be valid in a state of the infrastructure. For the malicious insider, this condition
has been the ``tipping point'' for an actor's state (given here as the parameterized \texttt{as a}).
To integrate insiderness to unintentional insiders, we simply add \texttt{unawareness} as an additional
sufficient condition to the rule.
\begin{ttbox}
Insider a C as \ttequiv tipping_point (as a) \ttor unaware (as a)
               \ttimp (\ttforall b \ttin C. UasI a b)
\end{ttbox}
Although the above insider predicate is a rule, it is not axiomatized. 
It is just an Isabelle definition, that is, it serves as an abbreviation. 
To use it in an application, like the auction protocol, we can use 
this rule as a local assumption in theorems 
or using the \texttt{assumes} feature of locales \cite{kpw:99}).

Based on the state transition and the above defined 
\texttt{sn\_scenario}, we define the first Kripke structure.
\begin{ttbox}
 sn_Kripke \ttequiv 
  Kripke \{ I. sn_scenario \ttrelIstar I \} \{sn_scenario\}
\end{ttbox}

\subsection{Attack: Eve can get data}
\label{sec:getatt}
How do we find attacks? The key is to use invalidation \cite{kp:14}
of the security property we want to achieve, here the global policy.
Since we consider a predicate transformer semantics, we use
sets of states to represent properties. 
The invalidated global policy is given by the following set \texttt{ssn}.
\begin{ttbox}
 ssn \ttequiv \{x. \ttneg (global_policy x ''Eve'')\}
\end{ttbox}
The attack we are interested in is to see whether for the scenario
\begin{ttbox}
 sn\_scenario \ttequiv  Infrastructure ex_graph local_policies 
\end{ttbox}
from the initial state \texttt{Isn \ttequiv \{sn\_scenario\}},
the critical state \texttt{ssn} can be reached,
that is, is there a valid attack \texttt{(Isn,ssn)}?

For the Kripke structure
\begin{ttbox}
 sn_Kripke \ttequiv Kripke \{ I. sn_scenario \ttrelIstar I \} Isn
\end{ttbox}
we first derive a valid and-attack using the attack tree proof calculus.
\begin{ttbox}
  \ttvdash [\ttcalN{(Isn,SN)}, \ttcalN{(SN,ssn)}]\ttattand{\texttt{(Isn,ssn)}}
\end{ttbox}
The set \texttt{SN} is an intermediate state where \texttt{Alice} moves to instagram
 to then put her data \texttt{''Alice's\_diary''} there.

The attack tree calculus \cite{kam:18b} exhibits that an attack is possible.
\begin{ttbox}
 sn_Kripke \ttvdash {\sf EF} ssn
\end{ttbox}
The attack tree formalisation in the Isabelle Infrastructure framework provides
adequacy, that is, Correctness and Completeness theorem for the relationship between
attack trees and the CTL statement \cite{kam:18b}.
We can thus simply apply the Correctness theorem \texttt{AT\_EF} to 
immediately prove CTL-{\sf EF} statements. This application of the meta-theorem 
of Correctness of attack trees saves us proving the CTL formula tediously 
by exploring the state space in Isabelle proofs. Alternatively, we could use 
generated code for the function \texttt{is\_attack\_tree} in Scala \cite{kam:21}
to check that a refined attack of the above is valid.

\section{Conclusions}
%
%
%
%
\subsection{Related work on awareness}
Awareness contributes to having knowledge of something; thus, security awareness could be considered as
a cognitive behavioural response to security and understanding its consequences. Some studies investigate
this possible understanding of internet and cyber security awareness, such as Bulgur \cite{bcb:10}.
Korovessis et al. \cite{kfph:17} introduces a ``toolkit approach to information security awareness and education'',
whilst focusing on organisations and the importance of user training by introducing a toolkit.
Training in this sense, focuses on teaching skills to safeguard information.
They completed a string of surveys, focus groups and interviews with different
participant groups and ages to establish the effectiveness of the toolkit. Results showed that the prototype
was successful in establishing awareness, however limitations were shown through the delivery of the approach
as the kit was not accessible to everyone.
Kruger and Kearney \cite{kk:06} establish a model prototype for assessing informational security awareness. The model
focuses on knowledge, attitude, and behaviour.
As stated by Lacey \cite{lac:09}, the gap in internet security is not the technology, but fundamentally the
awareness in people. The effectiveness of the approach is assessed by the resulting attitudes and behaviour to the topic. 

Bada, Sasse and Nurse \cite{bsn:15} also investigated through a psychological perspective where lack of
motivation lead to poorly designed security systems and poor security compliance.
The study results showed a
raised awareness and had positive effects on creating a ``security minded culture''. By introducing human
factors to awareness campaigns, the results deemed more positive, showing us that security awareness can
be increased if the tool used is more personal and relatable.
Bada, Sasse and Nurse \cite{bsn:15} provide a literature based survey on the effectiveness of campaigns
on human behaviour comparing cyber security awareness campaigns in Africa and UK.
They review Dolan et al's nine critical factors which influence and change human behaviour. Although
these factors provide an even finer granularity of categorizing human motivations, they are aligned
with the psychological characterization by Homans \cite{hom:61} that we we use as a basis for our model.
While our work uses an experimental approach, their survey \cite{bsn:15} also leads them to conclude that
``security education has to be more than providing information to users – it needs to be targeted,
actionable, doable and provide feedback''. Our approach is aligned with their findings, since our security
tool and modeling enables a ``targetted, actionable, doable'' analysis  of a social network leading
to feedback to the user.

Labuschagne et al \cite{lbve:11}
proposed a game hosted by social networking sites to increase security awareness.
The game uses social networks, something that is accessible by those at home and at work.
Lack of security knowledge is what makes people vulnerable and unable to protect their information,
an idea clearly stated by Kritzinger and von Solms \cite{ks:10}, with internet becoming so involved
in personal lives, it is paramount that the tools to raise awareness should be accessible to all. The
approach utilises a medium that is popular, therefore accessible. Whilst a prototype has not been
created, the approach must be analysed to see if it would utilise the increase of public awareness
to internet security. In this scenario the game that is hosted by social media sites is an approach
that would possibly be interacted with by the younger audience, producing a limitation as to its
non-inclusive medium, leaving a numerous amount of the public not being educated to.
Jemal Abwajy 2012 [18] concludes that combined delivery methods of text, video and game would be a more
suitable approach to deliver security awareness, rather than individual as it creates an inclusive
audience.

\subsection{Related work on Isabelle Insider and Infrastructure framework}
\label{sec:relisa}
A whole range of publications have documented the development of the Isabelle Insider framework.
The publications \cite{kp:13,kp:14,kp:16} first define the fundamental notions of insiderness, policies,
and behaviour showing how these concepts are able to express the classical insider threat patterns
identified in the seminal CERT guide on insider threats \cite{cmt:12}.
This Isabelle Insider framework has been applied to auction protocols \cite{kkp:16,kkp:16a}. An Airplane
case study \cite{kk:16,kk:21} revealed the need for dynamic state verification leading to 
the extension of adding a mutable state. Meanwhile, the embedding of Kripke structures and CTL
into Isabelle have enabled the emulation of Modelchecking and to provide a semantics for attack
trees \cite{kam:17a,kam:17b,kam:17c,kam:18b,kam:19a}.
Attack trees have provided the leverage to integrate Isabelle formal reasoning for IoT systems
as has been illustrated in the the CHIST-ERA project SUCCESS \cite{suc:16} where 
attack trees have been used in combination with  the Behaviour Interaction Priority (BIP) component 
architecture model to develop security and privacy enhanced IoT solutions.
This development has emphasized the technical rather than the psychological side of the framework
development and thus branched off the development of the Isabelle {\it Insider} framework into the
Isabelle {\it Infrastructure} framework. Since the strong expressiveness of Isabelle allows to formalize
the IoT scenarios as well as actors and policies, the latter framework can also be applied to 
evaluate IoT scenarios with respect to policies like the European data privacy regulation
GDPR \cite{kam:18a}. Application to security protocols first pioneered in the
auction protocol application \cite{kkp:16,kkp:16a} has further motivated the analysis of Quantum Cryptography
which in turn necessitated the extension by probabilities \cite{kam:19b,kam:19c,kam:19d}.

Requirements raised by these various security and privacy case studies have shown the need for a
cyclic engineering process for developing specifications and refining them towards implementations.
A first case study takes the IoT healthcare application and exemplifies a step-by-step
refinement interspersed with attack analysis using attack trees to increase privacy by ultimately
introducing a blockchain for access control \cite{kam:19a}. This formalisation of secure distributed data
labels has given rise generalising to sets of blockchain for Inter-clockchain protocols \cite{kn:20}. 
First ideas to support a dedicated security refinement process are available in a preliminary
arxive paper \cite{kam:20a} but the first to fully formalize the RR-cycle
and illustrate its application completely is the application to the Corona-virus Warn App (CWA) \cite{kl:20}.

\subsection{Discussion and Outlook}
We have presented a pragmatic action research study into awareness in social networks. User awareness 
interviews have given evidence to design, implement, and test a web-based tool enabling to show the user
how much is known about her. This feedback leads users to be more cautious and not give private data to
social networks. In our research, we have followed the action research methodology, that is, we have used
quantitative and qualitative research with practical interventions which consisted in implementing
a web application tool based on social network APIs for feedbacking to users what is visible of their data. 
In addition, we mechanized formal modeling and analysis for social network scenarios including
human actors in Isabelle.
For the latter application, we have used the Isabelle Insider framework to provide a dynamic logic model
enabling 
(1) formally reproducing the experimental scenario and (2) embedding the notion of awareness in the general
security notion of insiderness. We have thus linked up social network analysis to formal security engineering
and provided a novel formal notion of unintentional insiderness.

\bibliographystyle{abbrv}
\bibliography{../../insider}
\end{document}